\documentclass[aps,prl,preprint,superscriptaddress,amsmath,amssymb,onecolumn]{revtex4-2}
\usepackage{graphicx}% Include figure files
\usepackage{dcolumn}% Align table columns on decimal point
\usepackage{bm}% bold math
\usepackage{multirow}
\usepackage{lineno}
\usepackage[colorlinks,linkcolor=blue,anchorcolor=blue,citecolor=blue] {hyperref}
\begin{document}
%Title of paper
\title{Topological Classical Systems with Generalized Chiral Symmetry}

\author{Zhang-Zhao Yang}
\thanks{These authors have contributed equally to this work.}
\author{An-Yang Guan}
\thanks{These authors have contributed equally to this work.}
\affiliation{Key Laboratory of Modern Acoustics, MOE, Institute of Acoustics, Department of Physics, Collaborative Innovation Center of Advanced Microstructures, Nanjing University, Nanjing 210093, People’s Republic of China}

\author{Xin-Ye Zou}
\email{xyzou@nju.edu.cn}
\author{Jian-Chun Cheng}
\email{jccheng@nju.edu.cn}
\affiliation{Key Laboratory of Modern Acoustics, MOE, Institute of Acoustics, Department of Physics, Collaborative Innovation Center of Advanced Microstructures, Nanjing University, Nanjing 210093, People’s Republic of China}
\affiliation{State Key Laboratory of Acoustics, Chinese Academy of Sciences, Beijing 100190, People’s Republic of China}

\begin{abstract}
The bulk band topology of symmetry invariant adiabatic systems in the thermodynamic limit are considered to be determined by the hopping energy. In this work, we present that in closed classical systems, due to generalized chiral symmetry broken, the on-site energy cannot always be regarded as identical and can crucially impact the topological properties of the systems. Based on a finite one-dimensional chain, we demonstrate that the non-equivalent on-site energy of bulk lattices affects the topological phases of the bands, and the on-site energy of end lattices affects the existence of the topological states. Along these lines, the correspondence with generalized chiral symmetry in acoustic system is rigorously proposed. Our work provides a new degree of freedom for topological classical systems, and can be generalized to higher-dimensions and non-Hermitian conditions.
\end{abstract}

\maketitle

% Introduction part
Since there is no limitation imposed by the Fermi level, the classical wave systems have been considered as promising platforms for studying various topological effects that originate from the condensed matter physics \cite{TI1,TI2,TI3,TI4}. By imitating the arrangement as well as the properties of the atoms with low energy, several classes of the theoretically predicted topological insulators (TIs) -- the quantum Hall insulators \cite{QH5,QH6,QH7}, the quantum spin Hall insulators \cite{QSH8,QSH9,QSH10} and the topological crystalline insulators \cite{TCI11,Slager1,Slager2,TCI12,TCI13,TCI14,TCI15,TCI18}, have been successfully realized in the optical \cite{OP19,OP20,OP21,OP22,OP23,OP24,OP25,OP26,OP27,OP28,OP29,OP30,OP31,OP32,OP33,OP34,OP35,OP36,OP37}, mechanical \cite{Mec37,Mec39,Mec40}, microwave \cite{MW41}, electric circuit \cite{CC42,CC43} and acoustic systems \cite{Mec38,AC44,AC45,AC46,AC47,AC48,AC49,YZZ,AC50,AC51,AC52,AC53,AC54,AC55,AC56,OP370} in multiple dimensions. 

On the one hand, in the thermodynamic limit, the time-reversal invariant theoretical models are considered naturally preserving generalized chiral symmetry, which guarantees the impacts of the on-site energy to be negligible. On the other hand, the reported works on realizing the topological effects in real physical systems always construct surfaces interacting with ``environment'' consisting of trivial domains. When symmetries are slightly reduced, the adiabaticity condition guarantees that additional effects can be regarded as perturbations applied on the on-site terms of the Hamiltonian. In such cases, the equivalent on-site energy does not affect the generation of the topological effects, and the topological phases are crucially determined by the hopping terms within and between the lattices. However, for closed finite structures, i.e., crystalline systems, the broken symmetry at boundaries can crucially affect the properties of the atoms in the corresponding lattices. In this case, the thermodynamic limit is not necessarily satisfied, and generalized chiral symmetry can be broken. Therefore, one natural question to consider is, what are the impacts of the on-site energy on the topological effects in systems with reduced symmetries? Only a few works have presented that the boundary conditions can crucially affect the existence of the topological states \cite{SAP57,SAP58,SAP59,SAP60}, and even defining the appropriate classical wave structures that conform the theoretical models still needs to be addressed.

In this work, we challenge the notion that the classical systems can always be characterized by Hamiltonians with generalized chiral symmetry, and the topological phases only depend on the hopping energy. Instead, we demonstrate that the boundaries of the closed systems always results in generalized chiral symmetry broken, and the nonequivalent on-site energy can strongly affect the topological properties. By judiciously adding additional diagonal terms to the Hamiltonian of the one-dimensional (1D) Su-Schrieffer-Heeger-like (SSH-like) chain with nonidentical atoms in one lattice, we present the impacts of the on-site energy on the topological phase and the generation of the topological states in closed systems. Crucially, we further demonstrate that these additional terms correspond exactly to the boundary conditions, i.e., the hard boundary and the soft boundary, in classical wave systems. Due to the universality of these results, the topological effects can be rigorously predicted.

% Section II

\begin{figure}[htbp]
	\centering
	\includegraphics[width=0.95\textwidth]{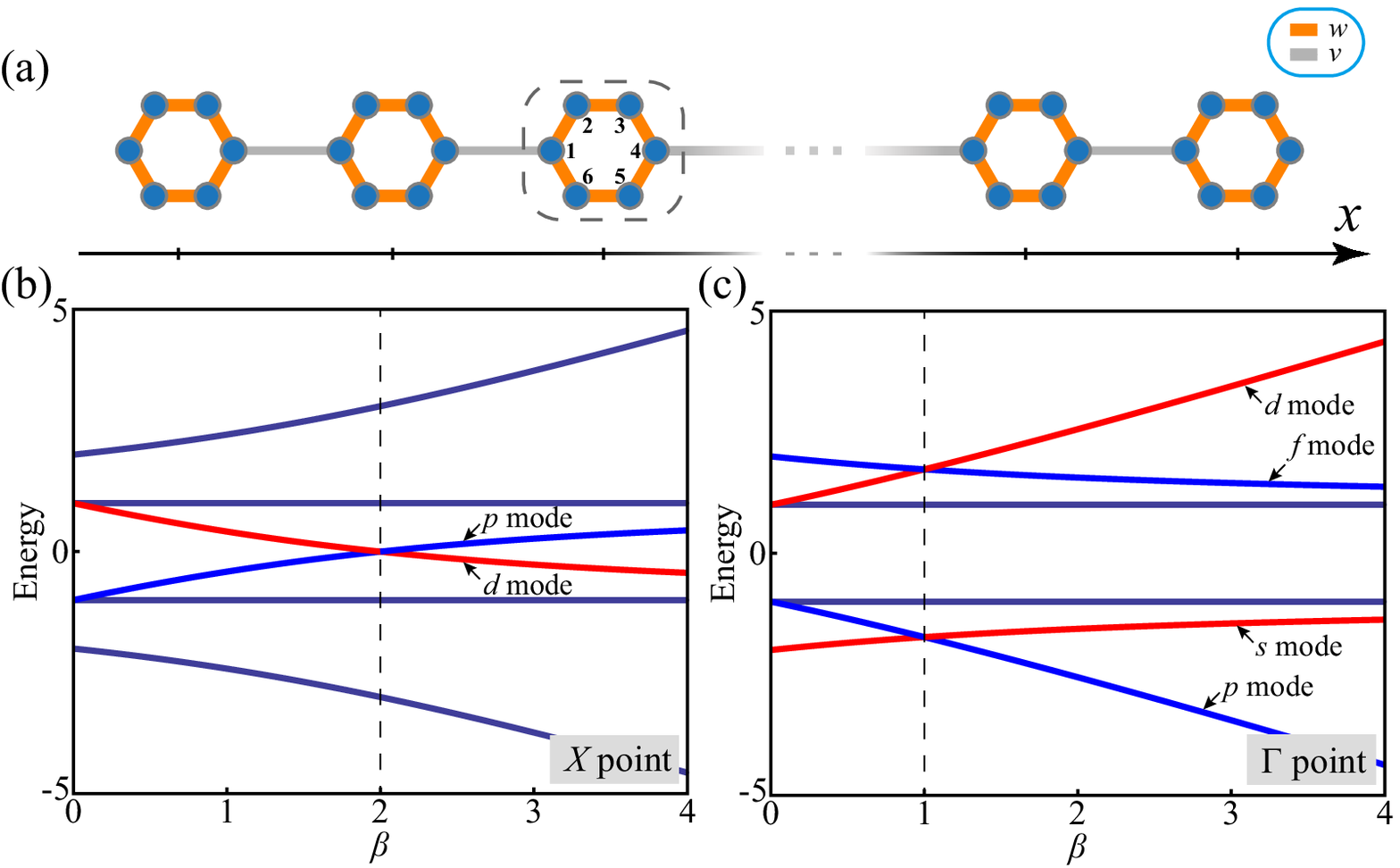}
	\caption{1D SSH-like model with generalized chiral symmetry. (a) Schematic of the chain. (b)-(c) The energy of the bands at (b) $ \Gamma $ point and (c) \textit{X} point varies as $ \beta $.}
	\label{figure1}
\end{figure}

We start from the 1D SSH-like model in the thermodynamic limit as shown in Fig. \hyperref[figure1]{1(a)}. Distinct from the SSH model, there are six atoms (labeled with 1 to 6, respectively) arranged in a honeycomb structure in the lattice. The intra- and inter- lattice hoppings are defined as $ w $ and $ v $, respectively. Here, we consider the zero on-site terms condition and the corresponding Hamiltonian of the lattice $ H_{0}(k) $ in $ k $-space can be written as
\begin{equation}
H_{0}(k)=-\sum_{\langle i;n\rangle}(wc_{i,n}^{\dagger}c_{i,n+1}+ve^{ik}c_{i,1}^{\dagger}c_{i,4}+{\rm H.c.}),
\label{eq1}
\end{equation}
where $ c_{i,n}^{\dagger} $ and $ c_{i,n} $ are the creation and annihilation operators of the $ n $-th atom in the $ i $-th lattice, respectively. It is obvious to see that, although lacking $ C_{6} $ symmetry, the preservation of inversion symmetry still guarantees that the geometric phase of the bands can be characterized by the 1D Zak phase
$ \mathcal{Z}_{n}=-\oint dk\langle u_{n}(k)\lvert i\partial_{k}\rvert u_{n}(k)\rangle $ \cite{Zak61}, where $ \rvert u_{n}(k)\rangle $ represents the Bloch wave function of the $ n $-th band. Therefore, $ \mathcal{Z}_{n}=0 $ indicates the ordinary band, and $ \mathcal{Z}_{n}=\pi $ indicates the topological band. By applying inversion symmetry, we then can obtain \cite{Zak62}
\begin{equation}
e^{i\mathcal{Z}_{n}}=\frac{\eta_{n}(X)}{\eta_{n}(\Gamma)},
\label{eq2}
\end{equation}
where $ \eta_{n}(P)$ $ (P = X, \Gamma) $ represents the $ \pi $-rotational even or odd parity of $ \rvert u_{n}(k)\rangle $ at the high symmetry point $ P $ of the first Brillouin zone. For conventions, the six Bloch wave function can be named as $ s $ mode (even parity), a pair of $ p $ mode (odd parity), a pair of $ d $ mode (even parity) and $ f $ mode (odd parity). As a result, the topological phase transition is determined by the corresponding parity inversion, which can be realized by tuning the hopping ratio $ \beta=v/w $.

Figures \hyperref[figure1]{1(b)} and \hyperref[figure1]{1(c)} show the corresponding band inversion spectra at $ X $ and $ \Gamma $ when $ w=1 $, respectively. It is noticed that the band inversion point at $ X $ corresponds to $ \beta=1 $, and the band inversion point at $ \Gamma $ corresponds to $ \beta=2 $, respectively. Note that there are two flat resonance band that can be excluded, and the topological property of a bandgap is determined by the summation of the Zak phases of all the bands below this gap \cite{Zak63}. As a result, the topological end states are predicted to emerge in the first and third gaps when $ \beta>1 $, and the topological end states in the second gap are predicted to emerge as `zero-energy' modes when $ \beta>2 $ (see Supplemental Material \cite{SM}, Sec. I). Figures. \hyperref[figure2]{2(a)-2(c)} present the energy band structures of the lattice when $ \beta=0.5 $, $ \beta=1.5 $ and $ \beta=3 $, respectively. The energy spectrum for the closed chain spanning 20 lattices is shown in Fig. \hyperref[figure2]{2(d)}. One essential feature of theoretical models is the equivalent on-site energy, which guarantees that the topological properties are determined by the hopping energy. However, we argue that for classical systems with reduced symmetries at boundaries, the impacts of the on-site energy cannot be neglected.

% Section III
We further consider the flow of the wave function in the system. Here, we define the impedances of the intra- and inter- lattice as $ Z_{w} $ and $ Z_{v} $, respectively. The intrinsic impedance of the atoms is defined as $ Z_{c} $. Due to $ C_{2} $ symmetry and inversion symmetry, there are two types of identical atoms in one lattice, which are labeled with 1, 4 and 2, 3, 5, 6, respectively [Fig. \hyperref[figure1]{1(a)}]. Therefore, for $ i $-th lattice, the flow of the wave function can be written as
\begin{equation}
-Z_{c}^{-1}\lvert \textbf{u}_{i}\rangle = A\lvert \textbf{u}_{i}\rangle
+ B\lvert \textbf{u}_{i-1}\rangle + C\lvert \textbf{u}_{i+1}\rangle,
\label{eq3}
\end{equation}
where
\begin{equation}
A = \begin{pmatrix}
2Z_{w}^{-1}+Z_{v}^{-1} & -Z_{w}^{-1} & 0 & 0 & 0 & -Z_{w}^{-1} \\
-Z_{w}^{-1} & 2Z_{w}^{-1} & -Z_{w}^{-1} & 0 & 0 & 0 \\
0 & -Z_{w}^{-1} & 2Z_{w}^{-1} & -Z_{w}^{-1} & 0 & 0 \\
0 & 0 & -Z_{w}^{-1} & 2Z_{w}^{-1}+Z_{v}^{-1} & -Z_{w}^{-1} & 0 \\
0 & 0 & 0 & -Z_{w}^{-1} & 2Z_{w}^{-1} & -Z_{w}^{-1} \\
-Z_{w}^{-1} & 0 & 0 & 0 & -Z_{w}^{-1} & 2Z_{w}^{-1} \\
\end{pmatrix},
\label{eq4}
\end{equation}
$ B_{14}=-Z_{v}^{-1} $, $ C_{41}=-Z_{v}^{-1} $ and $ \lvert \textbf{u}_{i}\rangle=[u_{1}, u_{2}, u_{3}, u_{4}, u_{5}, u_{6}]^{\rm T} $. Next, we define the parameters $ w=-\omega^{2}Z_{c}/Z_{w} $ and $ v=-\omega^{2}Z_{c}/Z_{v} $. By replacing the impedances with $ w $, $ v $ and $ \omega^{2} $, Eq. \hyperref[eq3]{(3)} then can be rewritten as $ \omega^{2}\lvert \textbf{u}_{i}\rangle = A_{w}\lvert \textbf{u}_{i}\rangle+B_{v}\lvert \textbf{u}_{i-1}\rangle+C_{v}\lvert \textbf{u}_{i+1}\rangle $, where $ A_{w} $ , $ B_{v} $ and $ C_{v} $ are the new transformed matrixes. After Fourier expansion, Eq. \hyperref[eq3]{(3)} finally can be rewritten as
\begin{equation}
\omega^{2}\lvert \textbf{u}_{i}\rangle=H_{1}\lvert \textbf{u}_{i}\rangle,
\label{eq5}
\end{equation}
where
\begin{equation}
H_{1}=\begin{pmatrix}
2w+v & -w & 0 & -ve^{ik} & 0 & -w \\
-w & 2w & -w & 0 & 0 & 0 \\
0 & -w & 2w & -w & 0 & 0 \\
-ve^{-ik} & 0 & -w & 2w+v & -w & 0 \\
0 & 0 & 0 & -w & 2w & -w \\
-w & 0 & 0 & 0 & -w & 2w \\
\end{pmatrix}.
\label{eq6}
\end{equation}
It is noticed that the on-site terms are not equivalent and $ Tr(H_{1}-2w\mathbb{I}_{6\times 6})\neq 0 $, which represents that chiral symmetry and generalized chiral symmetry is broken (see Supplemental Material \cite{SM}, Sec. II). Meanwhile, compared with $ 2w $, the additional diagonal term $ v $ that originates from $ C_{2} $ symmetry cannot be regarded as perturbation. By solving $ H_{1} $ when $ w=1 $, the corresponding band inversion spectra at $ X $ and $ \Gamma $ are presented in Figs. \hyperref[figure3]{3(a)} and \hyperref[figure3]{3(b)}. It is obvious to see that only the third and the fourth band at $ X $ are inverted when $ \beta=2/3 $, and there is no inversion at $ \Gamma $. Accordingly, the energy structure of the nontrivial lattice when $ \beta=1 $ and the energy spectrum for the closed chain spanning 20 lattices are shown in Figs. \hyperref[figure3]{3(c)} and \hyperref[figure3]{3(d)}, respectively. It is obvious to see that the chiral-symmetry-protected topological states in the first and second gaps all vanish. Therefore, in addition to the hopping energy,  generalized chiral symmetry broken can crucially affect the topological phase of the systems.

% Section IV
We argue that in addition to the nontrivial bulk band topology, generalized chiral symmetry broken in the end lattices can also crucially affect the existence of the topological states in closed systems. In the following, instead of considering the shapes of the structure, we focus on discussing the impacts of the on-site energy in the end lattices. As depicted in Fig. \hyperref[figure4]{4(a)}, we now study the 1D closed SSH-like chain spanning 20 lattices when $ w=1 $ and $ \beta=3 $, and the on-site energy of all the lattices are full-identical as $ 2w+v $, which guarantees the three gaps of the system to be topological. Thus, symmetry-protected topological states are predicted to emerge in the end lattices. Further, we consider the cases in which symmetry of the end lattices are reduced [illustrated in Fig. \hyperref[figure4]{4(a)}]. The diagonal terms at the first atom in the first lattice and the sixth atom in the last lattice are replaced with $ 2w+\alpha v $, where $ \alpha $ is a correction factor. Figure \hyperref[figure4]{4(b)} shows the change of the topological states with different $ \alpha $. It is obvious to see that the topological states deviate along with $ \alpha $. When $ \alpha<0 $ or $ \alpha>2 $, the topological states degenerate with the bulk modes and are hard to be identified. Note that $ \alpha=0 $ and $ \alpha = 2 $ represent that generalized chiral symmetry in the end lattices is completely broken; $ \alpha=1 $ represents that symmetry is preserved, which actually corresponds to the thermodynamic limit. The energy spectra for $ \alpha=0 $ and $ \alpha=1 $ are presented in Figs. \hyperref[figure4]{4(c)} and \hyperref[figure4]{4(d)}, respectively. As a result, the on-site energy of the end lattices can also crucially affect the existence of the topological states in nontrivial structures, and considering large classical systems preserving generalized chiral symmetry is not always appropriate.

Based on the results above, we present a concise method to control the generalized chiral symmetry in topological systems. In our model, an additional term $ D\lvert \textbf{u}_{i}\rangle $ can be judiciously added into Eq. \hyperref[eq3]{(3)}, and the equation can be rewritten as $ -Z_{c}^{-1}\lvert \textbf{u}_{i}\rangle = A\lvert \textbf{u}_{i}\rangle + B\lvert \textbf{u}_{i-1}\rangle + C\lvert \textbf{u}_{i+1}\rangle + D\lvert \textbf{u}_{i}\rangle $, where
\begin{equation}
	D = \alpha Z_{v}^{-1}\begin{pmatrix}
		Q & O \\
		O & Q \\
		\end{pmatrix},
	\label{eq7}
\end{equation}
where $ Q_{22}=Q_{33}=1 $. Note that the additional term $ D\lvert \textbf{u}_{i}\rangle $ represents that wave function can go through an impedance $ Z_{v}/\alpha $ to zero, which can be considered as the extra boundary on the corresponding atom. Crucially, in classical wave systems, $ \alpha=0 $ actually correspond to the hard boundary conditions, and $ \alpha=1 $ corresponds to the soft boundary conditions (see Supplemental Material \cite{SM}, Sec. III). Notably, the tunable term $ D $ is decoupled with other parameters, which implies a novel degree of freedom. As a result, the $ D $ in the bulk lattice can control the bulk band topology, and the $ D $ in the end lattices can control the energy of the topological states.

% Section V
Finally, we propose an 1D closed acoustic chain as depicted in Fig. \hyperref[figure5]{5(a)} to demonstrate the discussions above. The finite structure consists of 20 periodic lattices, and the atoms are simulated by hexagonal resonators that are connected by tubes. Therefore, the intra- and inter- lattice hoppings can be marked as $ 1/l_{w} $ and$ 1/l_{v} $, respectively, where $ l_{w} $ and $ l_{v} $ are the corresponding lengths of the tubes [depicted in Fig. \hyperref[figure5]{5(a)}]. Note that in this system, the impedances $ Z_{c}=1/ i\omega C $, $ Z_{w}=i\omega L_{w} $ and $ Z_{v}=i\omega L_{v} $, where $ C $ is the acoustic capacitance of the resonators, $ L_{w} $ and $ L_{v} $ are the acoustic mass of the tubes, respectively. Here, we set the side length of the resonators $ r=1{\rm cm} $, the width of the tubes $ d=0.1{\rm cm} $, $ l_{w}=4{\rm cm} $ and $ l_{v}=0.6{\rm cm} $. Therefore, the intra- and inter- lattice hoppings can be calculated as $ w=8.81\times 10^{6} $ and $ v=20w/3 $. Meanwhile, we note that there is no need to construct the deep ``environment'' stacked by the trivial lattices to enclose the topological structure as the previous works. Instead, the finite surface of the finite structure analogs the open boundary of the closed condensed matter systems. When defining the boundaries of the bulk lattices as soft, the calculated eigenfrequences of the finite structure in the cases when the boundaries of the end lattices are hard or soft are presented in Figs. \hyperref[figure5]{5(b)} and \hyperref[figure5]{5(c)}, respectively. As a result, the topological end states emerge when the end boundaries are soft, and are close to the bulk modes when the end boundaries are hard. The pressure field distributions of the three topological states solved by finite-element method and the SSH-like model are depicted in Figs. \hyperref[figure5]{5(d)-5(f)}, respectively. 

% Conclution part
In summary, we have demonstrated the impacts of generalized chiral symmetry on the topological properties of the closed classical systems, and that the real physical systems are not always appropriately described by the theoretical models that are in the thermodynamic limit. We connect the on-site energy with the boundary conditions of the classical systems, which provides a possible platform to control the topological states to emerge at desired energy. In particular, the additional term $ D\lvert \textbf{u}_{i}\rangle $ proposed in this work is decoupled with the other parameters, which offers a novel degree freedom for designing topological systems, and greatly broadens applications of topological phenomena. We believe that our work not only paves a new way to study the topological effects in classical systems in different dimensions, but also can be generalized for non-Hermitian systems.

% Acknowledgments
\begin{acknowledgments}
This work was supported by the National Key R\&D Program of China (Grant No. 2017YFA0303700), National Natural Science Foundation of China (Grant Nos. 11634006, 11934009, and 12074184), the Natural Science Foundation of Jiangsu Province (Grant No. BK20191245), State Key Laboratory of Acoustics, Chinese Academy of Sciences.
	\end{acknowledgments}

% Reference
%\bibliography{References}
%

\newpage
% FIGURES
% Figure 1

% Figure 2
\begin{figure}[htbp]
	\centering
	\includegraphics[width=0.95\textwidth]{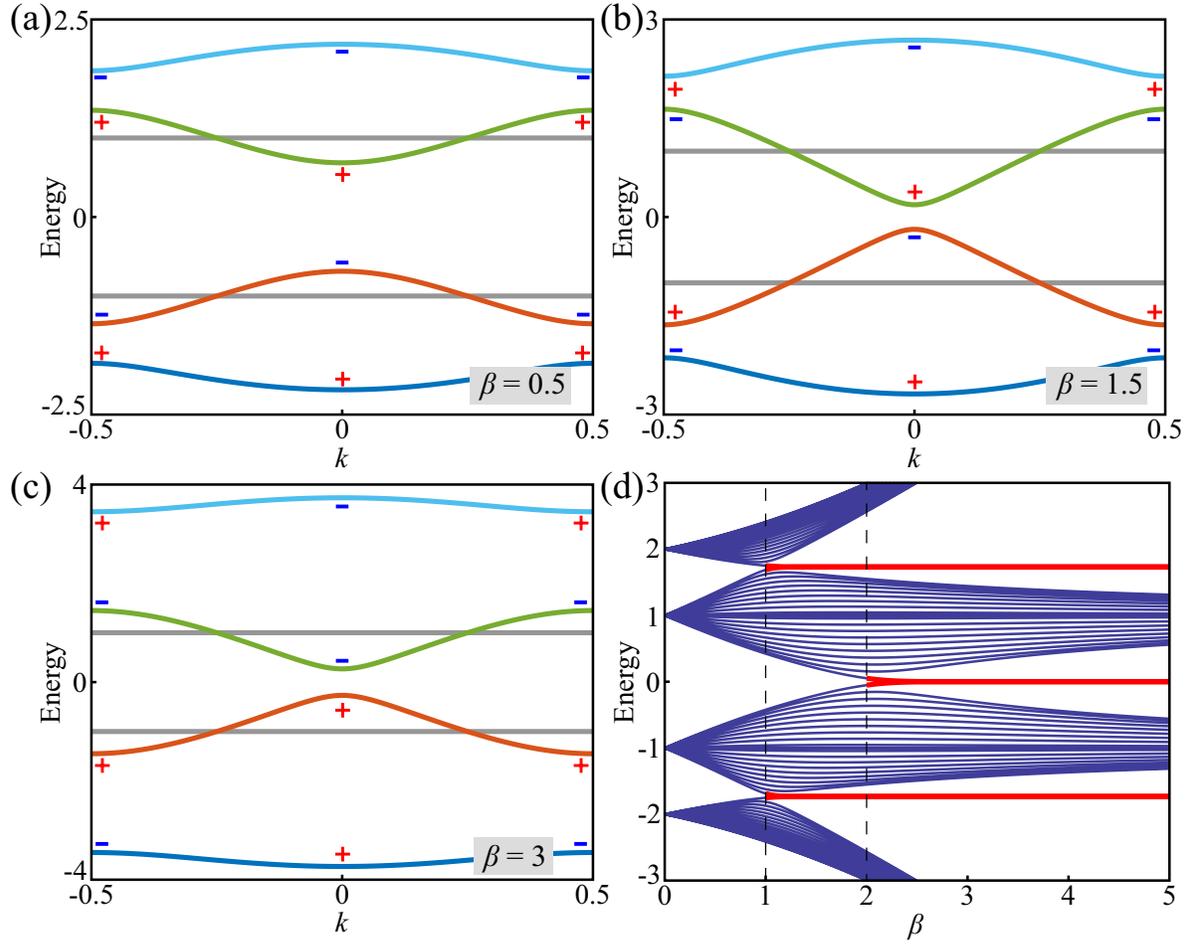}
	\caption{1D SSH-like model with generalized chiral symmetry. (a)-(c) Energy band structures of the lattice when (a) $ \beta=0.5 $, (b) $ \beta=1.5 $ and (c) $ \beta=3 $. (d) Energy spectrum of the finite chain with open boundaries.}
	\label{figure2}
\end{figure}

% Figure 3
\begin{figure}[htbp]
	\centering
	\includegraphics[width=0.95\textwidth]{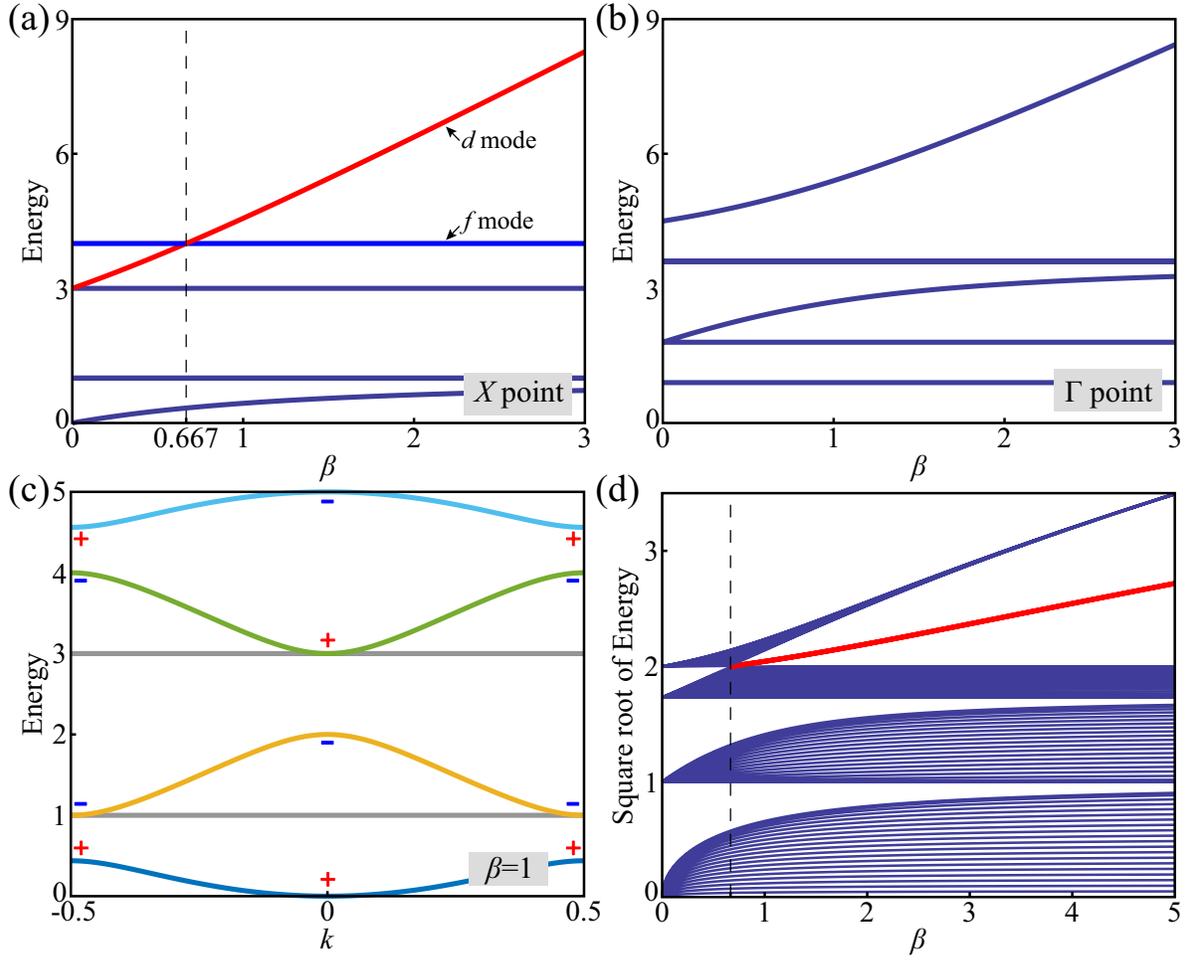}
	\caption{1D SSH-like model with generalized chiral symmetry broken. The energy of the bands at (a) $ \Gamma $ point and (b) \textit{X} point varies with $ \beta $. (c) Energy band structure of the lattice when $ \beta=1 $. (d) Energy spectrum of the finite chain with open boundaries. }
	\label{figure3}
\end{figure}

% Figure 4
\begin{figure}[htbp]
	\centering
	\includegraphics[width=0.95\textwidth]{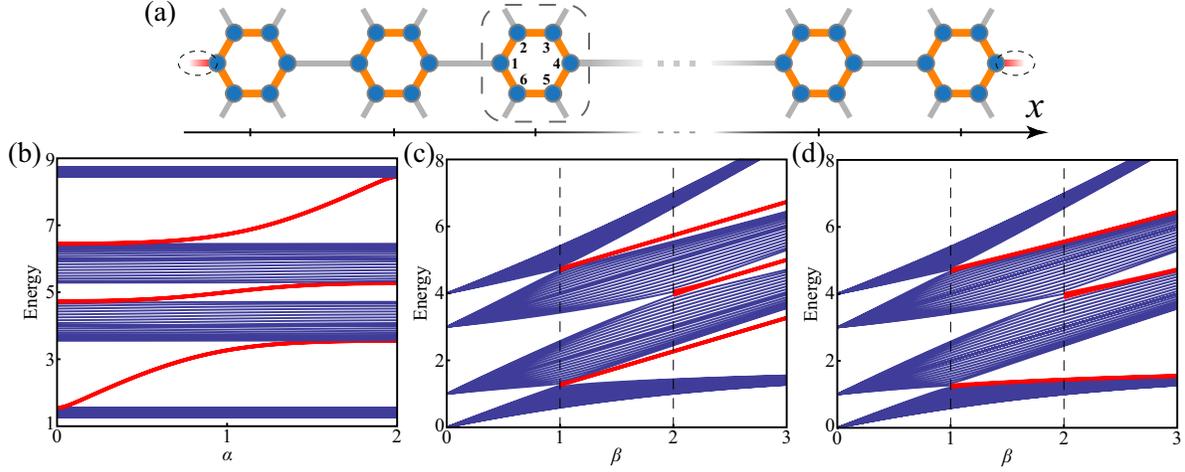}
	\caption{1D SSH-like model with generalized chiral symmetry broken in the end lattices. (a) Schematic of the finite chain. The gradual red lines at the ends indicates additional impedances. (b) The energy of the topological end states in the three nontrivial gaps varies with $ \alpha $ when $ \beta=3 $. Energy spectra when (c) $ \alpha=1 $ and (d) $ \alpha=0 $.}
	\label{figure4}
\end{figure}

% Figure 5
\begin{figure}[htbp]
	\centering
	\includegraphics[width=0.95\textwidth]{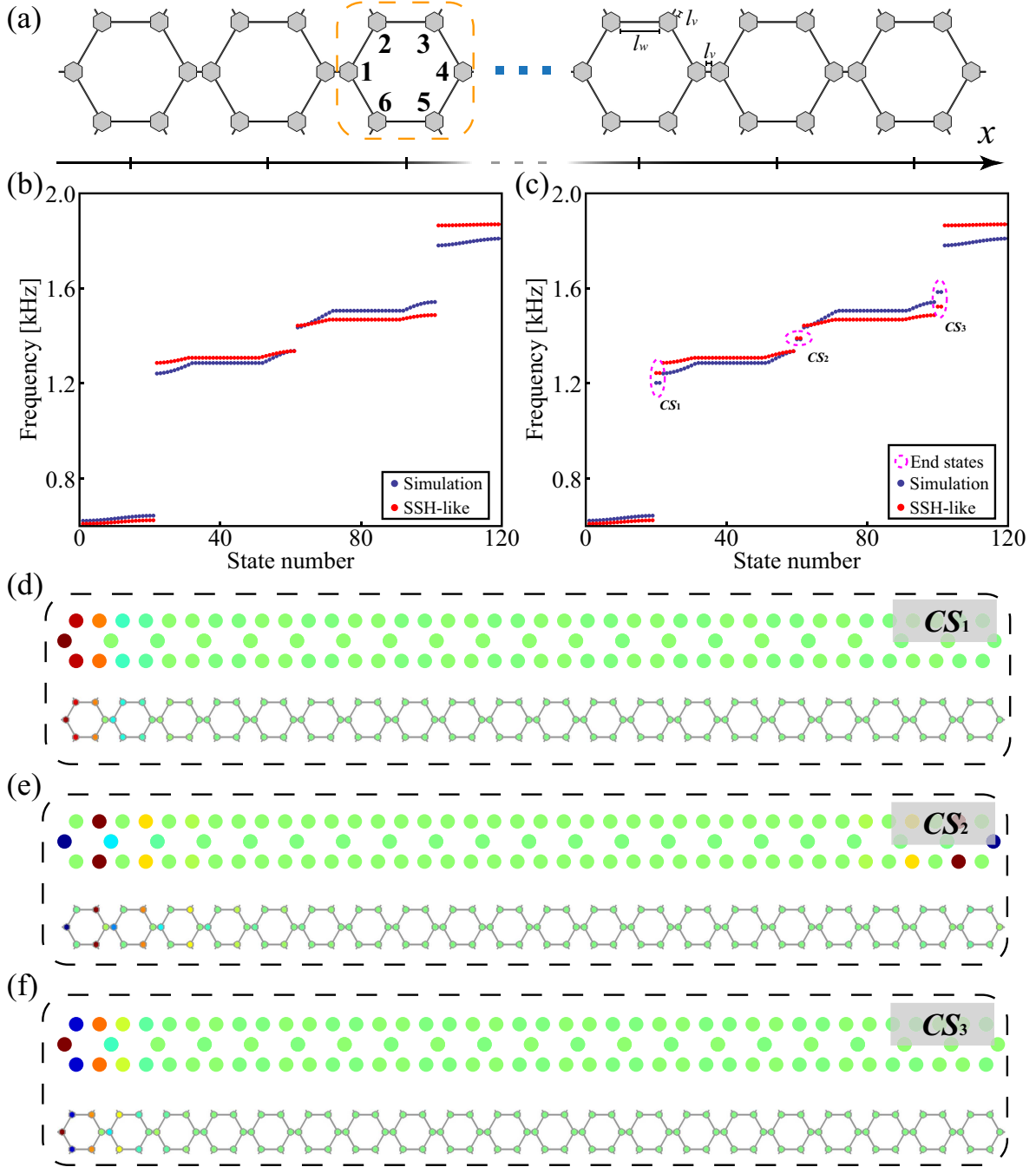}
	\caption{Acoustic topological chain. (a) Schematic of the 1D acoustic chain. (b)-(c) Calculated eigenfrequency spectra by using finite-element method (marked with blue) and SSH-like theoretical model (marked with red) when (b) hard boundary condition and (c) soft boundary condition. (d)-(f) Calculated field distributions of three types of topological states (labeled as $ \textit{CS}_{1} $, $ \textit{CS}_{2} $ and $ \textit{CS}_{3} $, respectively) by using finite-element method (bottom) and SSH-like theoretical model (top), respectively. }
	\label{figure5}
\end{figure}

\end{document}